\documentstyle[preprint,tighten,aps,multicol,epsfig]{revtex}
\begin{document}

\def\beq{\begin{equation}}
\def\eeq{\end{equation}}
\def\bea{\begin{eqnarray}}
\def\eea{\end{eqnarray}}
\def\beast{\begin{eqnarray*}}
\def\eeast{\end{eqnarray*}}
\def\lag{\langle}
\def\rag{\rangle}
\def\fnote#1#2{\begingroup\def\thefootnote{#1}\footnote{#2}
\addtocounter{footnote}{-1}\endgroup}

\draft
\preprint{KIAS-P99022, hep-ph/9903465}
\title{Neutrino Oscillations and R-parity Violating Collider Signals}

\author{S. Y. Choi, Eung Jin Chun, Sin Kyu  Kang, and Jae Sik Lee}

\address{School of Physics, Korea Institute for Advanced Study, 
         207-43 Cheongryangri-dong, Dongdaemun-gu, Seoul 130-012, Korea}

\maketitle

\begin{abstract}
R-parity and L
violation in the MSSM would be the origin of the neutrino oscillation 
observed in Super-Kamiokande. 
A distinctive feature of this framework is that it can be tested in colliders
by observing decay products of the destabilized LSP.
We examine all the possible decay processes of the neutralino LSP
assuming the bilinear contribution to neutrino masses dominates over the
trilinear one which gives rise to the solar neutrino mass.
We find that it is possible to probe neutrino oscillations
through colliders in most of the R-parity conserving MSSM parameter space.
\end{abstract}



\section{introduction}

Lepton (L) or baryon number (B) conservation is not a consequence
of gauge invariance in supersymmetric extension of the standard model.
The presence of both L and B violation would lead to fast proton decay,
and thus one usually imposes a discrete symmetry which ensures 
L and/or B conservation.  A typical example is R-parity discarding
renormalizable L and B violating operators, as a consequence of which the 
lightest supersymmetric particle (LSP) is stable and a good candidate of 
dark matter.
Another interesting possibility is to impose baryon parity (B-parity) which
allows the following L violating terms in the superpotential
of the minimal supersymmetric standard model (MSSM):
\beq \label{lvsup}
 \mu_i \hat{H}_u \hat{L}_i + 
 {1\over2}\lambda_{ijk} \hat{L}_i \hat{L}_j \hat{E}^c_k +
 \lambda'_{ijk} \hat{L}_i \hat{Q}_j \hat{D}^c_k \,.
\eeq
For the later use, it is convenient to generalize the bilinear term
to include the usual $\mu$--term: $\mu_\alpha \hat{H}_u \hat{L}_\alpha$ where
$\alpha=0,...,3$ and $\hat{L}_0=\hat{H}_d$.
It remains an open question what kind of discrete symmetry can be derived 
from a more fundamental theory beyond the MSSM.  
In view of gauge symmetry origin of discrete symmetries,
it has been shown that the standard R-parity and $Z_3$ B-parity can only be
consistent with the MSSM and the latter is preferred to arrange for
proton stability without fine-tuning taking into account higher dimensional
operators\cite{ibaros}.

R-parity (and L) violation in the MSSM would be the origin of 
neutrino masses and mixing, in particular,
the recently observed atmospheric neutrino oscillation \cite{sk-atm}.  
In the presence of the $|\Delta L|=1$ operators in the superpotential 
(\ref{lvsup})
as well as in the soft terms, neutrinos get nonzero masses through the
well-known mechanism of particle-sparticle mixing and one-loop radiative
correction \cite{hasu}.  A distinctive feature of this mechanism is 
that the LSP is destabilized and its decay modes can be easily identified
in colliders.  
The most characteristic events for those L-violating phenomena
are same-charge final-state dileptons due to the Majorana properties of two
produced LSP's, which have
little Standard Model backgrounds \cite{rpsig}.
Therefore, detecting the LSP decay, 
one can directly measure the quantities related to the L violating parameters 
\cite{viss,jaja},
which might enable us to extract some crucial information on the neutrino
sector of the theory and test neutrino oscillations in colliders.

The purpose of this paper is to provide a detailed analysis of 
the decay processes of the LSP (preferably taken as the lightest neutralino)
through both the bilinear and the trilinear L violating couplings in the MSSM 
whose sizes are determined by current neutrino experiments.
The bilinear terms play an important role in the whole discussion.
As they are typically the dominant sources for neutrino masses, 
our analysis will be based on the assumption that 
the atmospheric neutrino masses and mixing are mainly due to the bilinear 
terms.  The collider signatures of the bilinear terms have been studied
in Refs~\cite{valle1,valle2,bisset,roy} taking only one lepton generation and
the corresponding (tau) neutrino with a large mass $\sim$ MeV.
The correlation between collider signals and the atmospheric neutrino 
oscillation has been first analyzed in Ref.~\cite{viss} and 
also in \cite{jaja}.
In this  paper, we extend these works to 
include fully the three lepton generations and analyze both 
two and three body decays of the LSP.  The latter become important when the
trilinear couplings are introduced to explain the solar neutrino oscillation.

In the next section, we provide a comprehensive formulation of two and 
three body decay rates of neutralinos through both the bilinear and 
trilinear couplings.
In section III, 
we estimate the sizes of the R-parity and  L violating parameters 
for which the current neutrino data can be accommodated.
Then, we examine in section IV
the region of neutralino mass parameters for which 
the LSP decay and its branching modes can be measured.
An emphasis is put on how to separate out the information on the bilinear
terms from the contribution of the trilinear couplings which are present 
to give the masses and mixing for the solar neutrino oscillation.
Conclusion is given in section V.

\section{Particle-Sparticle Mixing and LSP Decays}

One usually rotates away the term $\hat{H}_u \hat{L}_i$ in Eq.~(\ref{lvsup}) 
so that L violation only appears in the trilinear couplings 
whose consequences in collider experiments are studied
extensively \cite{drmo}.   
But, there also exist soft supersymmetry breaking  bilinear terms which cannot
be rotated away in general.  These terms may play more important roles for
the LSP decay \cite{viss,valle1,valle2,bisset,roy}.
The relevant soft terms in the scalar potential are
\bea
V_{\rm soft} = m^2_{H_d} |H_d|^2 + m^2_{L_i} |L_i|^2 +
     \left\{ B H_u H_d + B_i H_u L_i + m^2_{L_iH_d} L_i H_d^\dagger 
   +\mbox{h.c.} \right\} + \cdots\,, 
\eea
from which it is clear that the L violating soft parameters 
$B_i, m^2_{L_iH_d}$ cannot be rotated away
unless the soft parameters satisfy certain relations called the alignment
condition \cite{bgnn}. 
This kind of alignment condition is usually achieved at the mediation scale
of supersymmetry breaking which is presumed to be L conserving.  However, 
at the weak scale where the minimization of the scalar potential takes
place, the alignment breaks down by the renormalization group evolution 
of the soft parameters. 
Therefore, unless one enforces the alignment at the weak scale
by some mechanism, the L violating terms being linear in $L_i$  
induce nonzero vacuum expectation values (VEVs) of sneutrino fields which
are non-aligned with $\mu_i$.  As a result of this misalignment, 
the mixing between particles and sparticles, 
in particular, neutrinos and neutralinos,
arises to lead important phenomenological consequences.

In the generic presence of the bilinear L violating terms $\mu_i$ and
nonzero sneutrino VEVs, the neutralino--neutrino mass matrix 
in the basis $(\tilde{B},\tilde{W}_3, \tilde{H}_u^0, \nu_\alpha)$
is given by 
\beq  \label{MN}
{\bf M_N}=
\pmatrix{ M_1 & 0 & M_Z s_W v_u/v  & -M_Z s_W v_\alpha/v \cr
           & M_2 & -M_Z c_W v_u/v  & M_Z c_W v_\alpha/v \cr
           &     &  0              & -\mu_\alpha \cr
           &     &                 &   0     \cr    }  \,.
\eeq
Here, $\bf M_N$ is a 7$\times$7 symmetric matrix,  
$v_{u}=\langle H_{u}\rangle$,
$v_0=\langle H_d \rangle$, 
$v_i=\langle \tilde{\nu}_i \rangle$ with $i=1..3$, 
and $s_W,c_W$ are the weak mixing elements.
The neutralino--neutrino mass matrix can be diagonalized by the rotation
$N^* {\bf M_N} N^\dagger$. That is, the index $i$ of the diagonalization 
matrix $N_{ij}$ runs for the mass eigenstates, and the index $j$ runs for
the weak eigenstates $(\tilde{B},\tilde{W}_3, \tilde{H}_u^0, \nu_\alpha)$.  
We would like to remind that for the mass eigenstates $\chi^0_i$,  
the states with $i=1,2,3$ are 
the neutrinos and $i=4$ denotes the lightest neutralino.

The chargino-charged lepton mass matrix in the basis $(\tilde{W}^+, 
\tilde{H}_u^+, E_k^+)$ and $(\tilde{W}^-,l_\alpha^-)$ is 
\beq
  {\bf M_C} = \pmatrix{ M_2 & \sqrt{2}M_Wv_\alpha/v  \cr
          \sqrt{2}M_Wv_u/v & \mu_\alpha \cr
              0          & M^l        \cr }\,,
\eeq
where $M^l$ is the 3$\times$4 matrix with the components; $M^l_{k0}=
-m^l_k v_k/v_0$, $M^l_{ki}= \delta_{ik} m^l_i$.
Two diagonalization matrices  $U_{R,L}$ are  defined by $U_R {\bf M_C} U_L^T$
where $U_R$ ($U_L$) casts the positively (negatively) charged states into 
the mass eigenstates.  The first index $i$ of the matrix 
$U_{Lij}$ and $U_{Rij}$ run for the mass eigenstates; 
the states with $i=1,2,3$ are
the charged leptons  ($e,\mu,\tau$)  and $i=4$ and 5 the two charginos.

In terms of the mass eigenstates, the neutral and charged current 
interactions take the form:
\beq \label{ffV}
 {\cal L}= {e\over 2s_W c_W} 
  \overline{\chi}^0_i \gamma^\mu \theta^N_{ij} P_L \chi^0_j Z_\mu
 - {e\over s_W} \left\{\overline{\chi}^0_i \gamma^\mu
   \left( \theta^L_{ij} P_L + \theta^R_{ij}  P_R \right) \chi^-_j
   W^+_\mu + {h.c.}  \right\}\,,
\eeq
where 
\bea \label{thetas}
 \theta^N_{ij} &=& \sum_{\sigma=3}^7\epsilon_\sigma N_{i\sigma}
      N^*_{j\sigma} \qquad\mbox{with}\qquad \epsilon_\sigma=(1, -1,....,-1)\,, 
 \nonumber\\
 \theta^L_{ij} &=& N_{i2}U^*_{Lj1}+ {1\over\sqrt{2}}
   \sum_{\sigma=4}^7 N_{i\sigma} U^*_{Lj\,\sigma\!-\!2}\,, \\
 \theta^R_{ij} &=& N^*_{i2}U^*_{Rj1}- {1\over\sqrt{2}} N^*_{i3}U^*_{Rj2}.
 \nonumber
\eea
To denote the mass eigenstates in a conventional way, let us introduce the 
following notations: $\nu_i= \chi^0_{i}$ with $i=1-3$ and 
$\tilde{\chi}^0_i=\chi^0_{i+3}$ for $i=1-4$.  

{}From Eq.~(\ref{ffV}), one gets the following rate of the decay process 
$\tilde{\chi}^0 \to \nu_i Z$ or $l_i^{\mp}W^{\pm}$ when the neutralino
$\tilde{\chi}^0$ is heavier than the $Z$ and/or $W$ boson:
\bea \label{twobody}
 \Gamma(\tilde{\chi}^0 \to  \nu_i Z )&=& {G_F M_{\tilde{\chi}^0}^3 \over 
       16\sqrt{2}\pi} I({M_Z^2 \over M_{\tilde{\chi}^0}^2})  
     \left| \theta^N_{\tilde{\chi} i} \right|^2\,,  \nonumber \\
\Gamma(\tilde{\chi}^0\to l^\mp_i W^\pm ) &=& {G_F M_{\tilde{\chi}^0}^3 \over 
       4\sqrt{2}\pi} I({M_W^2 \over M_{\tilde{\chi}^0}^2})\left\{ 
            \left| \theta^L_{\tilde{\chi} i} \right|^2 + 
   \left| \theta^R_{\tilde{\chi} i} \right|^2 \right\}\,,
\eea
where $I(x)=(1-x)^2(1+2x)$.  Note that we have neglected 
the possible neutralino decay into the lightest Higgs $h^0$ whose
rate is suppressed compared to the decay into $W$ boson by factor
of the order of $M_W^2/m_{{\tilde \chi}^0}^2$.  
There could exist other LSP decay modes such as
$\tilde{\chi}^0\rightarrow \nu\gamma$ which occurs at the one-loop level and
has the branching ratio at most a few percent \cite{muro}.

For the sake of discussion it is useful to use a seesaw formula valid for 
$v_i/v_0, \mu_i/\mu \ll 1$. This allows us to
decompose the diagonalization matrices into two parts; the usual neutralino
and chargino mixing, and the mixing between neutrinos (charged leptons) and
neutralino (charginos) which arise from L violation.
In this decomposition, the elements $\theta_{\tilde{\chi}i}$ for
the particle--sparticle mixing in Eq.~(\ref{thetas})
take the following factorized forms \cite{jaja}:
\bea \label{thetas'}
 \theta^N_{\tilde{\chi} i} &=& \left[ \sum_{a=1}^2 N_{\tilde{\chi} a} c^N_a 
    + 2 N_{\tilde{\chi} 3} c^N_{3}\right] \xi_i c_\beta\,, \nonumber \\
 \theta^L_{\tilde{\chi} i}  &=& \left[ N_{\tilde{\chi}2}c^L_1
       - {1\over\sqrt{2}} \sum_{a=1}^2 N_{\tilde{\chi} a} c^L_a 
       + {1\over\sqrt{2}} N_{\tilde{\chi} 4}(c^L_2 -c^N_4) \right] 
          \xi_i c_\beta \,,\\
 \theta^R_{\tilde{\chi} i} &=& N^*_{\tilde{\chi} 3} 
   \left[ c^R_1 - {1\over\sqrt{2}} c^R_2 \right] \xi_i c_\beta\,, \nonumber 
\eea
where $\xi_i\equiv v_i/v_0 - \mu_i/\mu$.  
Note that no mixing effect can
arise if the sneutrino VEVs are aligned with $\mu_i$, that is, $\xi_i=0$
as discussed in the beginning of this section.
The coefficients $c^N, c^L$ and $c^R$ are given by \cite{jaja,nowa}
\bea \label{c's}
&& c^N_i = {M_Z \over F_N} \left( -s_W {M_2 \over M_{\tilde{\gamma}} },
           c_W { M_1 \over  M_{\tilde{\gamma}} }, 
           -s_\beta {M_Z \over \mu}, 
           c_\beta {M_Z \over \mu} \right)\,, \nonumber\\
&& c^L_i = {M_W \over F_C} \left( \sqrt{2}, 2s_\beta {M_W \over \mu} \right)
   \,,\\
&& c^R_i = {M_W m^l_i \over F^2_C}
    \left( \sqrt{2}({M_2\over \mu} + {1\over t_\beta}), -{2\over c_\beta} 
       ({M_2^2 \over \mu M_W} + c^2_\beta{M_W \over \mu}) \right)\,, \nonumber
\eea
where $F_N = M_1 M_2/ M_{\tilde{\gamma}} - M_Z^2 s_{2\beta}/\mu$,
$F_C =  M_2 - M_W^2 s_{2\beta}/\mu$, and 
$ M_{\tilde{\gamma}}=c^2_W M_1 + s^2_W M_2$. 
In Eq.~(\ref{c's}),  $c^N_4$ and $c^L_2$ have 
extra $\mu_i/\mu$ contributions \cite{nowa} which are irrelevant
since they are canceled out in the physical process of neutralino decays 
as can be seen from the expressions in Eq.~(\ref{thetas'}).
Eqs.~(\ref{thetas'}) and (\ref{c's}) provide clear  
understanding of a few properties.   The ratios $\xi_i/\xi_j$
can be determined by measuring the  branching fractions of the decay modes
$\tilde{\chi}^0\to  l_i W$ independently of the other 
(L conserving) supersymmetric 
parameters like $M_2$ or $\mu$ also independently of the sizes of the trilinear 
couplings.  The rate $\Gamma(\tilde{\chi}^0\to lW)$ depends on
the charged lepton masses  through $c^R$  (\ref{c's}).  
This mass dependence is suppressed by $m^l_i/F_C$,
but can be significant for Higgsino--like LSP  and 
for large $\tan\beta$ as $c_2^R$ is proportional to $1/c_\beta$.

\medskip

When the LSP is lighter than $W$ boson, it has only three body decay 
modes which have two major contributions: 
one is the bilinear contribution coming
from the mixing $N, U_{R,L}$ through the exchange of $W$ or $Z$ boson and 
the other comes from the trilinear L violating couplings 
$\lambda,\lambda'$ through exchanges of squarks or sleptons. 
For the latter contributions, there exists a complete formulation 
in the literature \cite{gondo}.  
There are also contributions due to the charged and neutral Higgs 
exchanges which is neglected in our computation since they are generically 
smaller than the corresponding contributions from the $W,Z$ exchanges.  
More important one comes from the slepton-Higgs mixing 
\cite{valle1,hemp} whose size is determined by the bilinear parameters 
$\mu_i, B_i$ and $m^2_{L_i H_d}$. This mixing effect can be translated to 
the trilinear couplings in the mass basis, which is included in our analysis.

In order to calculate the rates of the neutralino 
decay $\tilde{\chi}^0\to\bar{f}_i \bar{f}_j f_k$,
we recast the decay amplitudes in terms of the 12 matrix elements into
the general form:
\bea \label{12M}
 {\cal M}( \tilde{\chi}^0 \to \bar{f}_i \bar{f}_j f_k) &=&
 S_{IJ}(s,t,u) [\bar{v}_{\tilde{\chi}} P_I v_i][\bar{u}_k P_J v_j]  \\ &+& 
 V_{IJ}(s,t,u) [\bar{v}_{\tilde{\chi}} \gamma^\mu P_I v_i]
               [\bar{u}_k \gamma_\mu P_J v_j]  \nonumber\\ &+&
 T_{IJ}(s,t,u) [\bar{v}_{\tilde{\chi}} \sigma^{\mu\nu} P_I v_i]
               [\bar{u}_k \sigma_{\mu\nu} P_J v_j]\,,  \nonumber
\eea
where $I,J=L,R$, and  $s=(p_{\tilde{\chi}}-p_i)^2$, 
$t=(p_{\tilde{\chi}}-p_j)^2$ and $u=(p_{\tilde{\chi}}-p_k)^2$.  
In our case, the neutralino decay  has the modes; 
$\tilde{\chi}^0 \to \nu \nu \bar{\nu}$, $\nu l \bar{l}$, $\nu q \bar{q}$ and 
$l q \bar{q}'$.   For each mode, nonvanishing matrix elements are presented
in Appendix.   The modes $\tilde{\chi}^0 \to \nu \nu \bar{\nu}$ and
$\nu u \bar{u}$ come solely from the bilinear contribution, and thus
has one and two matrix elements, respectively.
Note that the bilinear contributions appear in $S_{LR}$ or $V_{IJ}$, and 
only $S_{LL}$ and $T_{LL}$ terms can interfere with each other
neglecting the final state masses in our cases.  It turns out 
this interference is negligible in the cases we will consider.

The mode $\tilde{\chi}^0 \to \nu l \bar{l}$ contains 9 flavor modes of the
charged leptons.  Thus one could get information on the flavor structure 
of the trilinear couplings $\lambda_{ijk}$ by identifying dilepton signals with 
missing energy.  In a detector, 
the modes $\tilde{\chi}^0 \to \nu q \bar{q}$ and  $l q \bar{q}'$ 
can be seen as two jets with missing energy ($\nu j j$) and
charged lepton with two jets ($ljj$).  
In the case of three body decays of the LSP, the values $\xi_i$ could be 
measured by separating out the bilinear contribution in the $l_ijj$ decay
modes.
But  the three body LSP decay length from the bilinear contribution alone
is typically larger than a few meters.
The LSP could decay fast enough when there are relatively large trilinear 
couplings, in which case a large trilinear contribution to a decay mode 
can dominate over a bilinear contribution  making it impossible to
measure $\xi_i$.  

One of the most promising ways of measuring $\xi_i$ through various
L-violating channels is to look for same-charge dilepton events \cite{rpsig},
$l_i l_j + 2W$ or $l_i l_j + 4j$, from two LSP's produced in, e.g.,
$e^+e^-$ collisions. As will be discussed in the following sections, the recent
Super-Kamiokande data imply that one should observe comparable numbers of the
same-sign $(\mu\mu)$, $(\mu\tau)$, and $(\tau\tau)$ events \cite{jaja} except
for large $\tan\beta$ and small $|\mu |$.

\section{Neutrino Oscillations and L Violating Parameters}

Let us estimate the sizes of the R-parity and  L violating parameters 
for which the current neutrino data can be accommodated.
In particular, our discussion is based on the assumption that 
R-parity violation gives rise to the 
neutrino masses and mixing explaining the atmospheric and solar 
neutrino oscillations.

As is well known \cite{hasu,hemp,jono,cawh,smvi,nipo}, 
the mixing between neutrinos and neutralinos
gives rise to the so-called tree mass.
The neutrino--neutralino mass matrix ${\bf M_N}$ has two massless modes 
corresponding to the lightest neutrinos and the only massive neutrino 
which is clear from the following seesaw-reduced rank--one neutrino mass 
matrix; 
\beq \label{mtree}
 m^{\rm tree}_{ij} = {M_Z^2 \over F_N} c_\beta^2 \xi_i \xi_j\,. 
\eeq
Let us first suppose that the tree mass gives rise to the largest neutrino 
mass $m_{\nu_3}$ that is responsible for the atmospheric
neutrino oscillation, namely, 
$m_{\nu_3}= \sqrt{\Delta m^2_{\rm atm}} \approx 0.05$ eV \cite{sk-atm}.
Then the overall size of $\xi\equiv \sqrt{\sum_i \xi_i^2}$ 
is determined from Eq.~(\ref{mtree}):
\beq \label{xi-atm}
 \xi= 0.7\times10^{-6} { 1\over  c_\beta} \left(F_N\over M_Z\right)^{1\over2}
        \left( m_{\nu_3} \over 0.05 {\rm eV} \right)^{1\over2}  \,.
\eeq
Given the  mass matrix (\ref{mtree}), one can find the following expressions
for the oscillation amplitudes \cite{jaja}:
\bea \label{th-atm}
 \sin^22\theta^{\rm atm}_{\mu\tau} &= &
      4{\xi_2^2 \over \xi^2}{\xi_3^2 \over \xi^2}\,, \\
 \sin^22\theta^{\rm atm}_{e e\!\!\!/} &=&
      4{\xi_1^2\over\xi^2}(1-{\xi_1^2\over\xi^2}) \,. \nonumber
\eea
The first quantity is measured to be $\sin^22\theta^{\rm atm}_{\mu\tau} \sim 1$
in the Super-Kamiokande which implies $\xi_2 \approx \xi_3$ \cite{sk-atm}.  
The second one is measured
in the CHOOZ experiment for $\nu_e$ disappearance experiment \cite{chooz} 
from which one finds $\xi_1^2/\xi^2 < 0.05$  in the region
$\Delta m^2_{\rm atm} > 2\times 10^{-3}$ eV$^2$.   Therefore,
as realized in Ref.~\cite{jaja}, colliders may provide independent 
checks of the Super-Kamiokande and CHOOZ results by measuring the ratios 
$\xi_i/\xi$ through the LSP decays discussed in the previous section. 
In other words, combining the results from the neutrino experiment and 
collider signatures, one could {\it prove} R-parity violation
as a source of neutrino masses.  For this to happen, of course, 
the LSP has to decay inside a detector.

\medskip

When the LSP is heavier than the $W$ boson and thus the two body decays 
$\tilde{\chi}_1^0 \to l^{\pm} W^{\mp}$ are allowed kinematically,
the decay length is typically less than 10 $cm$ [see the next section].
For the LSP mass below $M_W$, the LSP can have only three body decay modes.
The LSP decay through $W,Z$ exchanges can make the decay length
smaller than 1 $m$ for the LSP mass close to $M_W$ 
even ignoring the effect of the trilinear couplings.
With larger trilinear couplings, the LSP decay through sfermion exchanges
can be detectable in colliders.  As will be seen in the next section,
the contribution of the trilinear couplings with $\lambda,\lambda' \sim
10^{-6}$ to the three body decay rates becomes comparable to 
the bilinear contribution in the case of common sfermion masses, 
$\tilde{m}_f = 300$ GeV.  
For the explicit calculation, we will take the {\it universal trilinear
couplings}.

The trilinear couplings of the order of $10^{-6}$ are too small to be 
influential
in neutrino experiments.  More interesting case is when the trilinear 
couplings are large enough to produce the neutrino masses and mixing
explaining the solar neutrino deficit \cite{smir}.
To estimate their sizes, 
let us recall that the trilinear couplings generate 
the one-loop neutrino mass through squark and slepton exchanges \cite{hasu}.
Two neutrinos which remain massless at tree level become massive
through the loop contribution.  The loop mass takes the form,
\bea \label{mloop}
 m^{\rm loop}_{ij} =  {3m_b^2 (A_b + \mu t_\beta) \over 
       8\pi^2  \tilde{m}_b^2 } \lambda'_{i33} \lambda'_{j33} +
   {m_\tau^2 (A_\tau + \mu t_\beta) \over 
      8\pi^2  \tilde{m}_\tau^2 } \lambda_{i33} \lambda_{j33}\,, 
\eea
where $A_{b,\tau}$ are trilinear soft parameters and $\tilde{m}_{b,\tau}$
are typical soft masses of sbottom and stau fields.
Here we keep only $\lambda_{i33}, \lambda'_{i33}$ assuming that 
they give the largest contribution which is the case when
the trilinear couplings follow the usual hierarchy of the Yukawa couplings, 
that is, the couplings involving the third generations are larger than 
the others.  This would be a consequence of a flavor symmetry dictating the
hierarchies among the Yukawa and trilinear couplings \cite{borz,chuns}. 
Still, having the arbitrariness of the L violating parameters $\xi_i, 
\lambda_{i33}$ and  $\lambda'_{i33}$, the loop mass can be made larger 
than the tree mass.  But in the context of minimal supergravity 
\cite{hemp,nipo,nar} or  gauge mediated supersymmetry breaking 
\cite{vemp,hwang,nelson} where the misalignment $\xi_i \neq 0$ is generated 
radiatively, the tree mass is typically larger than the loop mass.
This is our motivation to assume that the largest mass from Eq.~(\ref{mtree})
explains the atmospheric neutrino oscillation, and the second largest mass 
due to the loop correction accounts for the solar neutrino oscillation.
Suppose now that $\lambda_{233}$ gives the second largest eigenvalue
$m_{\nu_2}$, which will be the case with the model discussed below. 
Then, one needs to have 
$m_{\nu_2}= \sqrt{\Delta m^2_{\rm sol}} \approx 2\times10^{-3}$ eV 
for the MSW  solution of the solar neutrino
deficit \cite{smir}.
Therefore,  the size of $\lambda_{233}$ has to be
\beq \label{lam-sol}
 \lambda_{233} \approx 1.2\times10^{-4} \left(3\over t_\beta \right)^{1\over2}
  \left(100 {\rm GeV} \over |\mu| \right)^{1\over2}
  \left(m_{\nu_2} \over 2\times10^{-3} {\rm eV}\right)^{1\over2}\,,
\eeq
taking $\tilde{m}_\tau = 300$ GeV and ignoring the trilinear soft terms $A$.
Let us note for the later use that the required value of $\lambda_{233}$ 
becomes smaller for larger $\tan\beta$.  It is expected that $\lambda_{133}$ 
is smaller than $\lambda_{233}$ by one or two orders of magnitude in the case
of the favorable small mixing MSW solution \cite{smir}.
As discussed in the previous section, one can use 
the decay modes $\tilde{\chi}^0_1 \to \nu l_i l_j$ 
to get information on the flavor structure of $\lambda_{ijk}$.

With such large trilinear couplings (\ref{lam-sol}), the three body 
decay rate can become even larger than the two body decay rate.  
To figure out when this can happen, let us note that the two 
body decay length for $M_{\tilde{\chi}_1^0}\sim 150$ GeV is about 
0.1 $cm$ [see FIG. 1].  Then using the Dawson formula for the three 
body decay rate of the photino--like LSP \cite{dawson}, 
we find that the same decay length can be obtained when
a single large coupling takes the value, 
\beq \label{lam-con}
\lambda, \lambda' \sim 1\times10^{-4}
  \left( \tilde{m}_f \over 300 \,{\rm GeV} \right)^2
  \left( 150\,{\rm GeV} \over M_{\tilde{\chi}_1^0} \right)^{5\over2}
  \left( 0.1\, cm \over l \right)^{1\over2}\,,
\eeq
where $\tilde{m}_f$ is the mass of squark and $l\equiv 1/\Gamma$ 
denotes the LSP decay length.
We find the above value is in the range of the required value (\ref{lam-sol})
for small $\tan\beta$.
Therefore, the effect of trilinear couplings
could surpass that of bilinear couplings so that
it would be hard to separate out the information on $\xi_i$ 
alone.  Furthermore, the loop masses from the 
trilinear couplings $\lambda'_{i33}$ in Eq.~(\ref{mloop})
may affect the largest neutrino mass $m_{\nu_3}$.  For instance, taking
the value of $\lambda'_{i33}$ as in Eq.~(\ref{lam-sol}), one finds that
the loop mass is comparable to the tree mass (\ref{mtree}).  Therefore,
the expressions (\ref{mtree}), (\ref{th-atm}) should be altered to
include the loop mass as well.   In the context of supergravity where
nonaligned $\xi_i$ are generated from the trilinear couplings, the loop mass
can be comparable to the tree mass to account for both the atmospheric
and solar neutrino oscillations \cite{ckkl}.
In this case, it would be nontrivial to probe the mixing angle through 
L violating decays of the LSP unless all the trilinear couplings are known.
However, there is a class of models which are more predictive and 
do not possess the above difficulties. 
This is the so-called {\it bilinear model} \cite{valle1,hemp,jono,vemp,nelson}.

\medskip

The bilinear model does not have trilinear R-parity violating couplings.  
A classic example would be the minimal SU(5)
grand unified theory where generalized $\mu$ terms, 
$\mu_\alpha \hat{H}_u \hat{L}_\alpha$,
are allowed but the trilinear couplings cannot be present to ensure
proton longevity.  
As discussed in the previous section, in order to calculate the LSP decays,
one has to go to  the mass basis diagonalizing the  neutralino/neutrino,
chargino/charged lepton and slepton/Higgs mixing \cite{hemp}. 
This amounts to generating effectively L violating trilinear couplings.  
Not bothering with details about bilinear and scalar mass parameters, 
it is expected that 
the trilinear couplings take the values, e.g.,  $ \lambda'_{ijk} \approx  
\xi_i \delta_{jk} m^d_j/ \langle H_d \rangle $. The dominant couplings
are then
\bea \label{lam-bi}
 \lambda'_{i33} &\approx& 2\times10^{-8} {1\over c_\beta^2}
            \left(F_N \over M_Z\right)^{1\over2}\,,   \nonumber\\
 \lambda_{i33} &\approx&  7\times10^{-9} {1\over c_\beta^2}
            \left(F_N \over M_Z\right)^{1\over2}   \,.
\eea
As was discussed in the previous section, these couplings are negligible
in our discussion for low $\tan\beta$ but become important for 
large $\tan\beta$.  The other couplings are negligible for all $\tan\beta$
in view of our previous discussions.
An interesting feature is that the loop mass can account for the solar neutrino
deficit through the MSW solution for large $\tan\beta$.  
For instance, taking $\tan\beta=50$,
we get $\lambda_{i33} \approx 2\times 10^{-5}$ which
agrees with the value given in Eq.~(\ref{lam-sol}).
We note that, in the large $\tan\beta$ region, the value of $\xi$ 
in Eq.~(\ref{xi-atm}) becomes comparable to $\lambda_{233}$.
In this case, the loop mass matrix has a large contribution from 
$\lambda'_{i33}$ but it also takes the form $m^{\rm loop}_{ij} \propto
\xi_i \xi_j$ with the change of its overall size.  Thus, 
the relations in Eq.~(\ref{th-atm}) remain untouched.
As will be shown explicitly in the next section,
the induced trilinear couplings are large enough
to make the LSP decay visible in colliders, but not large enough  
to change the two body decay length.
Furthermore, the numbers $\xi_i$ can be measured without any contamination
from the trilinear couplings.

\section{Numerical analysis of LSP decays}

Before presenting our numerical results, let us recapitulate the crucial 
points for testing neutrino oscillations through the detection of 
the R-parity violating LSP decays  in colliders.
First of all, the LSP has to decay inside a detector. So, we require the 
LSP decay length $l$ to be less than 1 $m$ and show
the corresponding region of neutralino mass parameters for given L 
violating parameters and sfermion masses.
We also assume the unification relation
\beq \label{unirel}
M_1={5\over3} t_W^2M_2  \,,
\eeq
and restrict ourselves to the region satisfying the recent
LEP constraint on the mass of the lightest chargino:
$M_{\tilde{\chi}^\pm_1}>90$ GeV \cite{opal}.  These restrictions
rule out the region of $M_2$ and $\mu$ for $M_{\tilde{\chi}_1^0} 
\lesssim 50$ GeV and the possibility of the chargino LSP.

Since the mode $l_iW$ or $l_ijj$ for the two or three body decay case
is important to determine the parameter $\xi_i$
[see Eqs.~(\ref{thetas'},\ref{mtree})], we will present several 
figures showing the branching ratios of all possible decay modes 
on the $M_2-\mu$ plane, as well as with fixed $M_2$ and varying $\mu$.
In the definition of {\it e.g.} the  $l_ijj$  branching ratio, we include both
$l_i^\pm$.   The quantities $\xi_i$ are measured by the bilinear 
contribution to the decay processes we need to separate out  this from 
the trilinear contribution.   For this purpose, we will present  
the branching ratios coming from bilinear terms and trilinear terms 
separately.
To get an idea of when the trilinear contribution becomes visible,
we will show the decay length and branching ratios taking 
the universal trilinear couplings of $10^{-6}$.
As discussed in the previous section,
when the trilinear couplings are of the order $10^{-4}$ which 
may be required to produce the solar neutrino mass, the trilinear contribution
to the two body decay rate becomes comparable to the bilinear contribution.
In this case, it would be meaningless to separate out the two contributions
making it complicated to probe the neutrino oscillations through 
collider signals.  
However, it is possible to have a good separation between the bilinear
and trilinear contributions while generating both the atmospheric and solar
neutrino masses.  This is the case in the bilinear model with large 
$\tan\beta$, where the order of the magnitudes of the
induced trilinear couplings are given in Eq.~(\ref{lam-bi}).  
Therefore, taking the bilinear model as a specific example for 
the atmospheric and solar neutrino oscillations, 
we will show the decay length and branching ratios.

In our computations, 
we take the following input values as a reference:
\bea \label{inputs}
&& \xi_1 << \xi_2=\xi_3 ={\xi \over \sqrt{2}  }\,,      \\
&& A_{\tilde{f}} = m_{\tilde{f}} = 
   300\,{\rm GeV}\quad\mbox{for all sfermions}\,, \nonumber
\eea
while $\xi$ fulfills the relation (\ref{xi-atm}).
We vary $M_2$ and $\mu$ in the range of  $50 \sim 500$  GeV and  
$-300\sim 300$ GeV. Here, we take $\xi_2 = \xi_3$ reflecting the large mixing
angle of the atmospheric neutrino oscillation as in Eq.~(13).

\medskip

Let us now present our results in FIGs.~1--8.
In FIG. 1, we show the decay lengths 
(a) including only the LSP decays to $W,Z$ bosons or through $W,Z$ exchanges, 
(b) taking the universal trilinear couplings $\lambda, \lambda'=10^{-6}$,
and (c) considering the bilinear model, for $\tan\beta=3$. 
The thick lines in the $M_2-\mu$ planes 
correspond to the points for $M_{\tilde{\chi}_1^0}=M_W$.
The shaded region is excluded by the constraint on the lightest chargino mass.
One finds that there is a very small region  with the two body decay length 
larger than 1 $m$ (denoted by filled stars just above the thick line) for 
negative $\mu$.  It is clear from FIG. 1a and 1c that 
the trilinear couplings in the bilinear model (\ref{lam-bi}) are
negligible for the  small $\tan\beta$.  There is also a little change in
FIG. 1b.  

The situation changes for $\tan\beta=50$ as in FIG. 2.
The decay lengths become shorter than 10 $cm$ in the two body decay 
region.  The $M_2-\mu$ plane is {\it completely}
covered for the bilinear model as shown in FIG. 2c.
This is mainly due to large $\lambda'_{i33}$ as in Eq.~(\ref{lam-bi}).
Also more region is covered in FIG. 2b with $\lambda,\lambda' =10^{-6}$.

In FIG. 3, we present (a) the $lW$ branching ratio adding all charged leptons,
and (b) each branching ratio fixing $M_2=300$ GeV, in the case of 
$\tan\beta=3$.  We do not show the region with $M_{\tilde{\chi}_1^0}<M_Z$ 
for which the $\nu Z$ channel is forbidden.
As can be seen in FIG. 3a, there is a limited region (denoted by dots)
where the $lW$ mode is strongly suppressed, which corresponds to the point 
with $\mu=-140$ GeV in FIG. 3b.  This implies that a cancellation among
various terms in $\theta^{L,R}_{\tilde{\chi} i}$ in Eq.~(\ref{thetas}) 
takes place.
As we put $\xi_2=\xi_3$, the $\mu W$ (solid line) and $\tau W$ (dashed line) 
branching ratios exactly coincide for large $|\mu |$ 
and thus for large $M_{\tilde{\chi}_1^0}$ ($M_2=300$ GeV).
But for smaller $|\mu |$ the two ratios deviate from each other
which contradicts with the result of Ref.~\cite{jaja}.
This is a consequence of the charged lepton mass dependence of
$\theta^R_{\tilde{\chi} i}$  in Eqs.~(\ref{thetas'}) and (\ref{c's}) 
which becomes very important for the Higgsino--like LSP.  
Let us mention as a reference that the valley at $\mu=-140$ GeV corresponds
to the LSP mass $M_{\tilde{\chi}_1^0}=123$ GeV.
Our result is larger than that of Ref.~\cite{viss} 
in the $\nu Z$ decay rate by factor of 2, coming from the fact that
the $\Delta L=-1$ and $\Delta L=1$ modes are to be summed.   
The $eW$ branching ratio is negligibly small as we put $\xi_1<<\xi_{2,3}$.  
If $\xi_1$ is comparable to $\xi_{2,3}$, the curve for the $eW$ branching
ratio is in parallel with that of the $\mu W$ branching ratio with
the constant deviation by factor of $\xi_1^2/\xi_2^2$ as the 
dependence on $m_e$ or $m_\mu$ is negligible.
However, the universal property does not remain for the $eW$ and 
$\tau W$ modes especially for large $\tan\beta$.

The same set of the branching ratios as FIG. 3 are shown in 
FIG. 4 with $\tan\beta=50$.
The $lW$ branching fraction becomes larger than that with smaller 
$\tan\beta$ and it can be bigger than 40 \%.  From FIG. 4b, one finds 
that as $\tan\beta$ increases the deviation between $\mu W$ and $\tau W$ 
branching ratios greatly magnifies as discussed in section II.

FIG. 5 shows three body branching ratios for $\tilde{\chi}_1^0\to\nu\nu\nu$, 
$\nu ll$, $\nu j j$ and $l_i jj$ without trilinear couplings.  
$M_2$ is fixed to be 150 GeV.
For  $\tan\beta=3$ in FIG. 5a, the $\mu jj$ (solid line) and 
$\tau jj$ (dashed line) branching ratios appear to be the same.  
On the contrary, there is a big deviation between two branching ratios
for $\tan\beta=50$ and small $|\mu|$ as shown in FIG 5b.
This is also due to the large dependence of $\theta^R_{\tilde{\chi} i}$ on the 
lepton masses for the Higgsino--like LSP.

In FIG. 6,
three body branching ratios with universal trilinear 
couplings of $10^{-6}$ with $M_2=150$ GeV and $\tan\beta=3$ are shown
by separating (a) the bilinear  and (b) the trilinear contributions.
As there is almost no interference between the two contributions, 
the full branching ratios  can be obtained just by adding these two
contributions as in FIG 6c.
The trilinear contribution to the $ljj$ modes is shown to be
negligible for positive $\mu$ in FIG. 6b.  
As can be seen from FIG. 6d,  the bilinear contribution (solid line)
dominates over the trilinear contribution (dashed line) except for
$\mu \sim -100$ GeV.

For $\tan\beta=50$ in FIG. 7,  trilinear contribution becomes larger
but still the bilinear contribution to $l_i jj$ mode dominates.
One again sees the big deviation between $\mu jj$ and $\tau jj$ modes for
smaller $|\mu|$.

Finally, the three body branching ratios of the bilinear model
with $\tan\beta=50$ are shown in FIG.~8.  Here the trilinear contribution 
dominates but it comes mostly from the $\nu jj$  mode.  
This is because the coupling $\lambda'_{i33}$ do not contribute 
to the mode $l_i jj$ since it involves heavy top production. 
Therefore, one can directly measure the $l_i jj$ branching ratios from the 
bilinear contribution which have   
branching ratios of $1-6$ \% as shown  in FIG. 8c.
Another important property is that the $\nu l l$ branching ratio
from the trilinear contribution is about $1 \%$ and thus one can also
probe the flavor structure of the trilinear couplings $\lambda$.
As a consequence, collider experiments can provide an independent check of 
whether the bilinear model explains the solar 
and atmospheric neutrino oscillations.

\section{conclusions}

Recent observation of neutrino oscillation in Super-Kamiokande would
require physics beyond the standard model.  Supersymmetric extension
of the standard model allowing L (and thus R-parity) violation is one of
the well-motivated theoretical frameworks for nonzero neutrino masses and 
mixing.  That is, the current atmospheric and solar neutrino data may be
explained by R-parity violation in the MSSM.
A distinctive feature of this picture is that it could be tested in 
colliders through L violating signatures of the LSP decay.
In generating neutrino masses from R-parity violation, one has to take into
account both trilinear and bilinear terms in the superpotential and in the
soft breaking scalar potential.  Typically the bilinear contribution 
to neutrino masses through nonaligned sneutrino VEVs dominates over the
trilinear contribution.  Based on this aspect, we have analyzed collider 
signals for the L violating decays of the LSP (taken as the lightest 
neutralino) which has two body and/or three body decay channels depending 
on the LSP mass.  

The results of our analysis are shown in FIGs.~1--8 presenting
the decay lengths and branching ratios as functions of the supersymmetric 
parameters $M_2$ and $\mu$ under the assumption of gaugino mass unification.
When the trilinear L violating couplings are too small, there is some region of
the LSP mass where the LSP decay cannot be seen in a detector.
However, if the trilinear couplings are large enough 
to give rise to the solar neutrino
mass while the atmospheric neutrino mass comes from the bilinear parameters,
the LSP is shown to decay inside a detector for the whole R-parity conserving
MSSM parameter space restricted by the recent lower bound on 
the chargino mass $m_{\tilde{\chi}^-_1} >90$ GeV and the 
unification relation of gaugino masses.
As a specific example of realizing both the atmospheric and solar neutrino
oscillations, we have taken the bilinear model with large $\tan\beta$
which has no arbitrariness in the trilinear couplings.  In such a model,
the neutrino oscillation parameters measured in the atmospheric neutrino
experiments can be directly probed in colliders by measuring 
the branching ratios of the LSP decays into $l_i W$ or $l_i jj$.
This opens a possibility to test the assumption  of the R-parity violating
MSSM as the origin of neutrino masses and mixing.

In a more general case with arbitrary trilinear couplings,  it would be
impossible to directly test the neutrino oscillations in colliders.
But if future colliders measure the $l_i W$ or $l_i jj$ branching ratios 
it would provide a hint for or against a specific mechanism 
predicting the structures of the trilinear couplings and/or soft supersymmetry
breaking parameters.

\section{Appendix}

In order to express the amplitude of the decay $\chi^0
\to \bar{f}_i \bar{f}_j f_k$ in a compact form, we introduce the notations:
\bea
 g^{\chi}_f &=& {\sqrt{2} e\over s_W c_W}[s_W N^*_{\chi1} Y_f 
                 + c_W N^*_{\chi2} T^3_f]\,,  \\
 h^{\chi}_f &=&\cases{\displaystyle 
         N^*_{\chi4}{ e\over\sqrt{2} s_W } {m_f \over M_W c_\beta}
           \quad \mbox{for $T^3=-{1\over2}$ fermions}\,, \cr
    \displaystyle  N^*_{\chi3}{ e\over\sqrt{2} s_W } {m_f \over M_W s_\beta}
           \quad \mbox{for $T^3={1\over2}$ fermions}\,, \cr } \nonumber\\
 l^{\chi}_i &=& {e^2\over \sqrt{2}s^2_W } \theta^L_{\chi i}\,, \quad
 r^{\chi}_i = {e^2\over \sqrt{2}s^2_W } \theta^R_{\chi i} \,, \quad 
 n^{\chi}_{if}= -{e^2\over 2s^2_W c^2_W } \theta^N_{\chi i}
                     [T^3_f-s_W^2Q_f] \,,\nonumber\\ 
 P_V(x) &=& {1\over x-M_V^2}\quad\mbox{for}\quad V=Z,W,\; x=s,t,u\,, 
        \nonumber\\
 P_{\tilde{f}}(x) &=& \cases{  \displaystyle
     { g^\chi_f (x-m^2_{\tilde{f}_R}) +  h^\chi_f \Delta_{\tilde{f}} \over
      (x-m^2_{\tilde{f}_1})(x-m^2_{\tilde{f}_2})  }     
     &for $x=s,t$\,, \cr  \displaystyle
   (m^2_{\tilde{f}_R} \leftrightarrow m^2_{\tilde{f}_L})
     & for $x=u$\,, \cr}             \nonumber\\
 Q_{\tilde{f}}(x) &=& \cases{ \displaystyle
    { h^{\chi*}_f (x-m^2_{\tilde{f}_R}) + g^{\chi*}_f \Delta_{\tilde{f}} \over
     (x-m^2_{\tilde{f}_1})(x-m^2_{\tilde{f}_2})  }  
      &for $x=s,t$\,, \cr  \displaystyle
     (m^2_{\tilde{f}_R} \leftrightarrow m^2_{\tilde{f}_L})
      & for $x=u$\,, \cr}     \nonumber
\eea
Here $m^2_{\tilde{f}_1, \tilde{f}_2}$ are the eigenvalues of the
sfermion mass-squared matrix,
$$
 \pmatrix{ m^2_{\tilde{f}_L} & \Delta_{\tilde{f}} \cr
           \Delta_{\tilde{f}} &  m^2_{\tilde{f}_R} \cr } \,,
$$
where $\Delta_{\tilde{f}}= m_f (A_f+\mu t_\beta)$  or
$m_f (A_f+\mu/t_\beta)$ for $T^3={\mp}{1\over2}$ sfermions, respectively.

For the amplitude of the decay $\chi^0\to \bar{\nu}_i e^+_j e^-_k$ coming from
the bilinear contribution and the trilinear coupling $\lambda_{ijk}$, 
we find the  following 5 nonvanishing matrix elements:
\bea
 S^{\nu e e}_{LL} &=& \lambda_{ijk}\left[ P_{\tilde{\nu}_i}(s) -{1\over2}
        P_{\tilde{e}_j}(t) -{1\over2} P_{\tilde{e}^c_k}(u) \right]\,, \\
 S^{\nu e e}_{LR} &=& \delta_{ik} r^\chi_j P_W(t)\,, \nonumber\\
 V^{\nu e e}_{LL} &=& \delta_{jk} n^\chi_{i\,e_j} P_Z(s)  \nonumber
    + \delta_{ik} l^\chi_j P_W(t) +{1\over2}\lambda_{ijk} Q_{\tilde{e}_k}(u)\,,
    \\
 V^{\nu e e}_{LR} &=& -\delta_{jk} n^\chi_{i\,e^c_j} P_Z(s) \nonumber
    - {1\over2}\lambda_{ijk} Q_{\tilde{e}^c_j}(t)\,, \\
 T^{\nu e e}_{LL} &=& -{1\over8} \lambda_{ijk} \left[P_{\tilde{e}_j}(t) 
        - P_{\tilde{e}^c_k}(u) \right]\,. \nonumber
\eea
Similarly, there are 5 nonvanishing matrix elements for the decay 
$\chi^0\to e^+_i \bar{u}_j d_k$:
\bea
 S^{eud}_{LL} &=& -\lambda'_{ijk} \left[  P_{\tilde{e}_i}(s) -{1\over2} 
    P_{\tilde{u}_j}(t) -{1\over2}P_{\tilde{d}^c_k}(u) \right]\,, \\
 S^{eud}_{LR} &=& -\lambda'_{ijk} Q_{\tilde{e}^c_i}(s)\,,  \nonumber\\
 V^{eud}_{LL} &=& V_{jk}^*l^{\chi}_i P_W(s) -{1\over2}\lambda'_{ijk}
    Q_{\tilde{d}_k}(u)\,,         \nonumber\\
 V^{eud}_{RL} &=& V_{jk}^*r^{\chi}_i P_W(s)\,, \nonumber\\
 T^{eud}_{LL} &=& +{1\over8} \lambda'_{ijk} \left[P_{\tilde{u}_j}(t)
        - P_{\tilde{d}^c_k}(u) \right]\,. \nonumber
\eea
We find that the decay  $\chi^0 \to \bar{\nu}_i \bar{d}_j d_k$ has 4 nonzero
matrix elements:
\bea
 S^{\nu dd}_{LL} &=&  \lambda'_{ijk}\left[ P_{\tilde{\nu}_i}(s) -{1\over2}
    P_{\tilde{d}_j}(t) -{1\over2}P_{\tilde{d}^c_k}(u) \right]\,,  \\
 V^{\nu dd}_{LL} &=& \delta_{jk} n^\chi_{i\,d_j} P_Z(s) +
     {1\over2}\lambda'_{ijk} Q_{\tilde{d}_k}(u)\,, \nonumber\\
 V^{\nu dd}_{LR} &=& -\delta_{jk} n^\chi_{i\,d^c_k} P_Z(s) -
     {1\over2}\lambda'_{ijk} Q_{\tilde{d}^c_j}(t)\,, \nonumber\\
 T^{\nu dd}_{LL} &=& -{1\over8} \lambda'_{ijk} \left[P_{\tilde{d}_j}(t)
        - P_{\tilde{d}^c_k}(u) \right] \nonumber \,.
\eea
For the decay  $\chi^0 \to \bar{\nu}_i \bar{u}_j u_k$, we find that there are 
only two nonvanishing matrix elements:
\bea
 V^{\nu u u}_{LL} &=& \delta_{jk}n^\chi_{i u_j} P_Z(s)\,, \\
 V^{\nu u u}_{LR} &=& -\delta_{jk}n^\chi_{i u^c_j} P_Z(s) \nonumber\,.
\eea
Finally, the decay $\chi^0 \to \bar{\nu}_i \bar{\nu}_j \nu_k$ has only one
nonvanishing term:
\beq
 V^{\nu\nu\nu}_{LL} = \delta_{jk}n^\chi_{i \nu_j} P_Z(s) 
  +  \delta_{ik}n^\chi_{j \nu_i} P_Z(t)  \,.
\eeq
%


\newpage
\section*{Figure Captions}

FIG. 1.  
Two body and three body decay lengths of the LSP on the $M_2$--$\mu$ 
plane with
(a) the vanishing trilinear couplings,
(b) the universal trilinear couplings of $\lambda ,\lambda' =10^{-6}$,
and (c) the trilinear couplings given by the bilinear model when
$\tan\beta=3$. The shaded area is excluded by the restriction
$m_{\tilde{\chi}_1^{\pm}}>90$ GeV and
the two body decay is allowed only
in the above the thick solid line corresponding to $m_{\tilde{\chi}_1^0}=M_W$.
Different markers are used to denote the region with the decay length of 
$l<10^{-3}~m$ (dot),
$10^{-3}<l<10^{-2}~m$ (open circle),
$10^{-2}<l<10^{-1}~m$ (asterisk),
$10^{-1}<l<1~m$ (filled circle),
$l>1~m$ (filled star) and
$l<1~m$ (filled box).  \\

FIG. 2. 
The same as FIG.~1, but with $\tan\beta=50$.\\

FIG. 3. 
(a) Two body branching ratio of the LSP decay into $l W$
on the $M_2$--$\mu$ plane with $\tan\beta=3$.
Different markers are used to denote the region with the branching ratio of
$~ 0\sim 0.2$ (dot),
$0.2\sim 0.4$ (open circle),
$0.4\sim 0.6$ (asterisk),
$0.6\sim 0.8$ (filled star),
$0.8\sim 0.9$ (filled triangle) and
$0.9\sim 1$ (filled box).
(b) The same branching ratios when $M_2=300$ GeV.
In the shaded area, $M_{\tilde{\chi}^0_1}<M_Z$.
The four different lines correspond to the decay mode of
$\tilde{\chi}^0_1\rightarrow \mu W$ (solid line),
$\tilde{\chi}^0_1\rightarrow  \tau W$ (dashed line),
$\tilde{\chi}^0_1\rightarrow l W$ (dash-dotted line) and
$\tilde{\chi}^0_1\rightarrow \nu Z$ (dotted line), respectively.\\

FIG. 4. 
The same as FIG.~3, but with $\tan\beta=50$.\\

FIG. 5.  
(a) Three body branching ratios with the vanishing trilinear couplings
when $M_2=150$ GeV and $\tan\beta =3$.
The shaded area is excluded due to the constraint 
$M_{\tilde{\chi}^{\pm}_1}>90$ GeV.
The five lines correspond to the decay mode of
$\tilde{\chi}^0_1\rightarrow \mu jj$ (solid line),
$\tilde{\chi}^0_1\rightarrow \tau jj$ (dashed line),
$\tilde{\chi}^0_1\rightarrow \nu jj$ (dash-dotted line),
$\tilde{\chi}^0_1\rightarrow \nu\nu\nu$ (thick dash-dotted line), \,and
$\tilde{\chi}^0_1\rightarrow  \nu ll$ (dotted line).
(b) The same as (a), but with $\tan\beta =50$.\\

FIG. 6. 
Three body branching ratio with universal trilinear couplings of
$10^{-6}$ when $\tan\beta =3$ and $M_2=150$ GeV. Bilinear and trilinear
contributions are shown in (a) and (b), respectively, and the sum of these two
contributions is shown in (c).
The line convention for each decay mode in (a), (b) and (c)
is the same as in FIG.~5.
In (d), we show the branching ratio summed over the five decay modes.
The solid line denotes the bilinear contribution and
the dashed line the trilinear contribution.\\

FIG. 7. 
The same as FIG.~6, but with $\tan\beta=50$.\\

FIG. 8.  
The same as FIG.~7, but with
the trilinear couplings given by the bilinear model. \\

\begin{figure}[ht]
\hspace*{-1.0 truein}
\psfig{figure=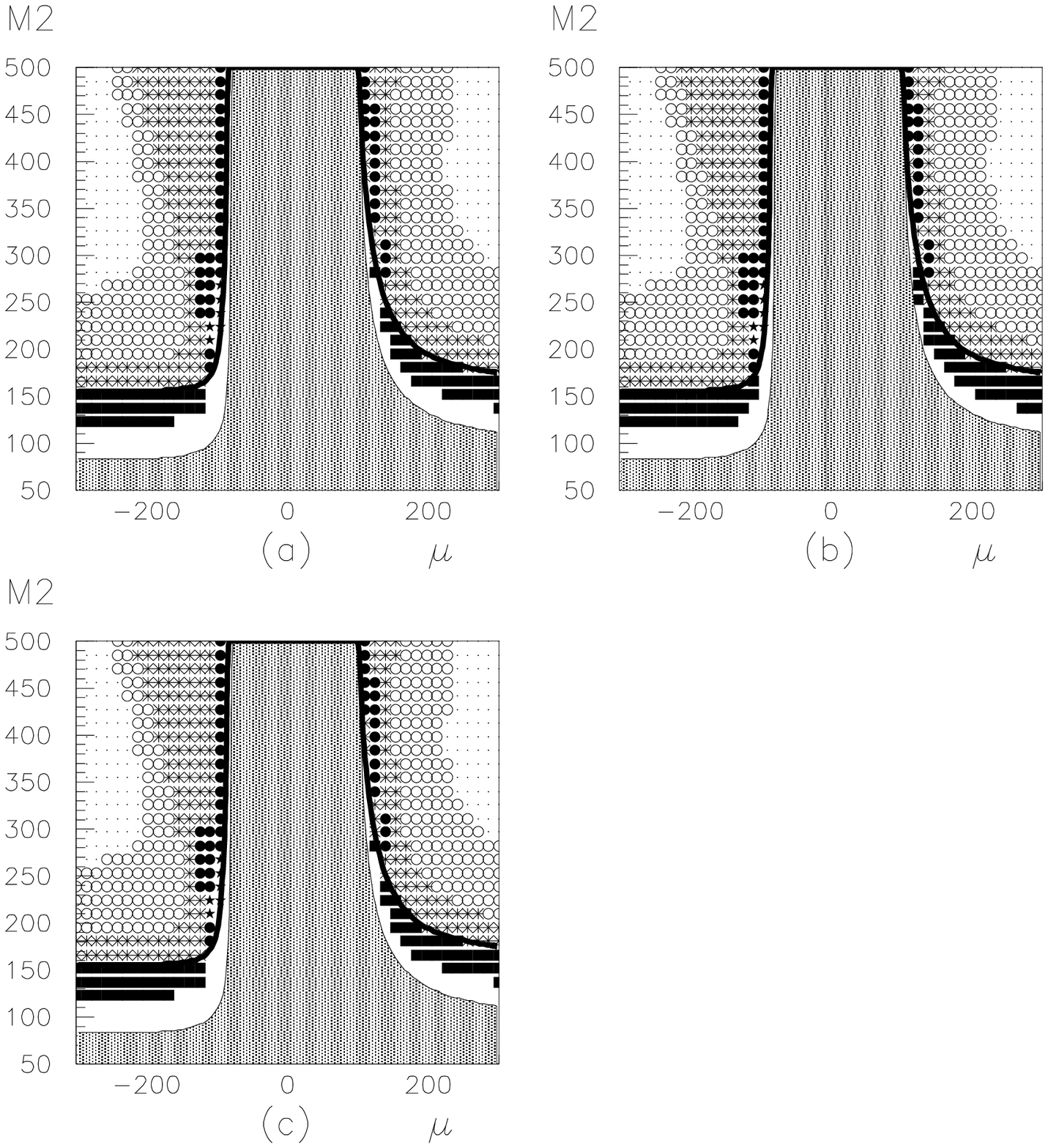}
\caption{}
\label{fig1}
\end{figure}

\begin{figure}[ht]
\hspace*{-1.0 truein}
\psfig{figure=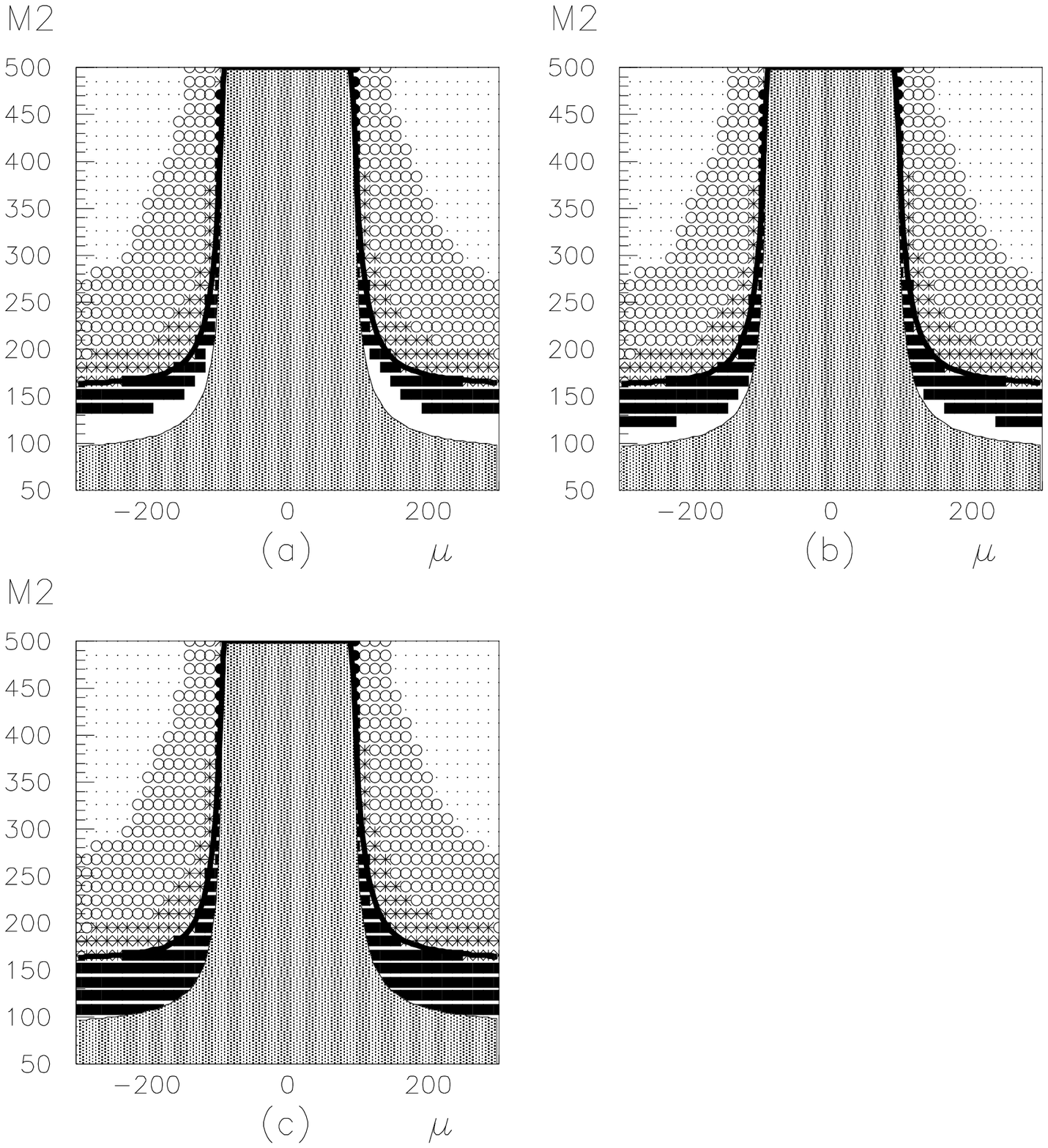}
\caption{}
\label{fig2}
\end{figure}

\begin{figure}[ht]
\hspace*{-1.0 truein}
\psfig{figure=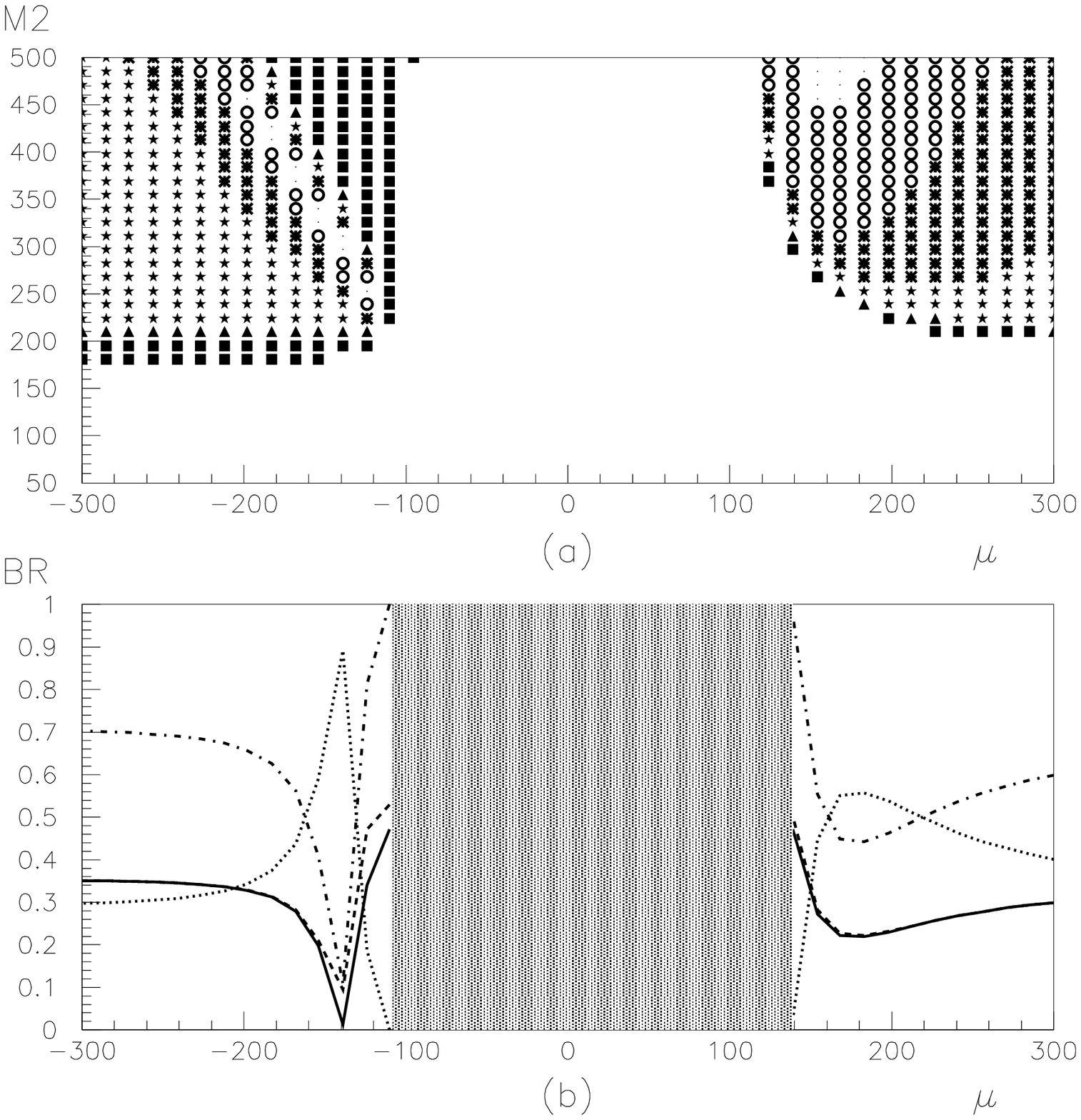}
\caption{}
\label{fig3}
\end{figure}

\begin{figure}[ht]
\hspace*{-1.0 truein}
\psfig{figure=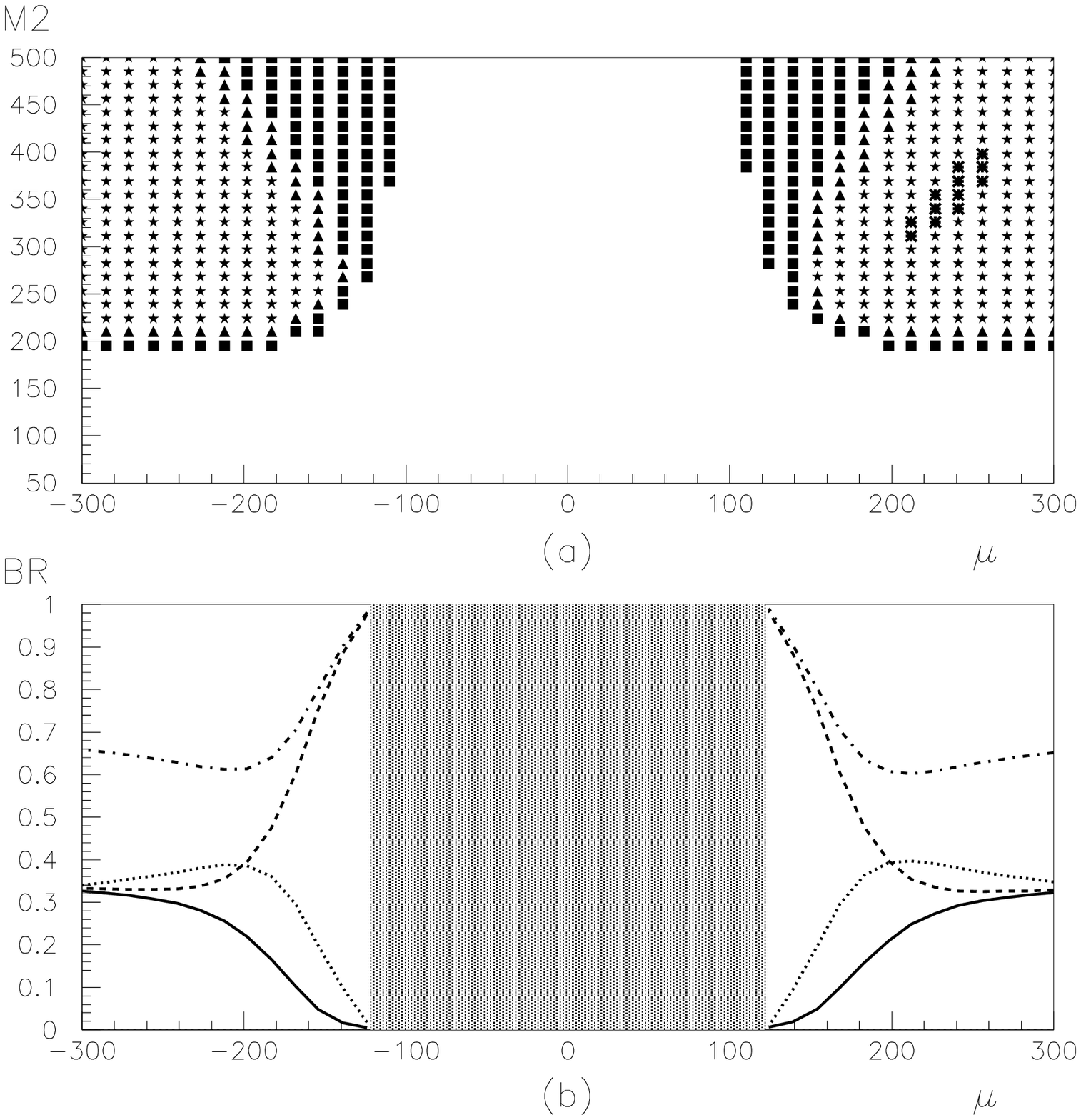}
\caption{}
\label{fig4}
\end{figure}

\begin{figure}[ht]
\hspace*{-1.0 truein}
\psfig{figure=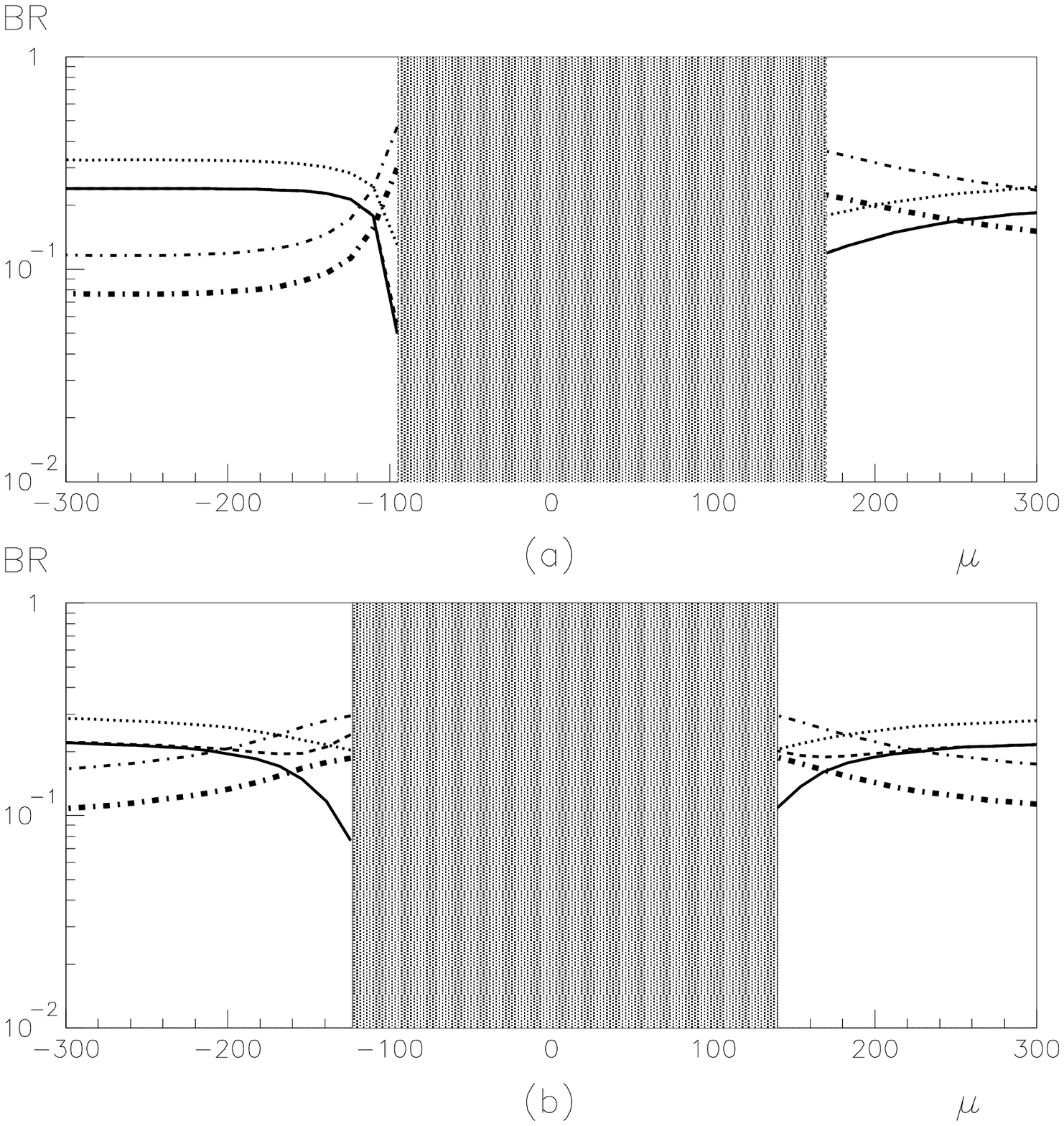}
\caption{}
\label{fig5}
\end{figure}

\begin{figure}[ht]
\hspace*{-1.0 truein}
\psfig{figure=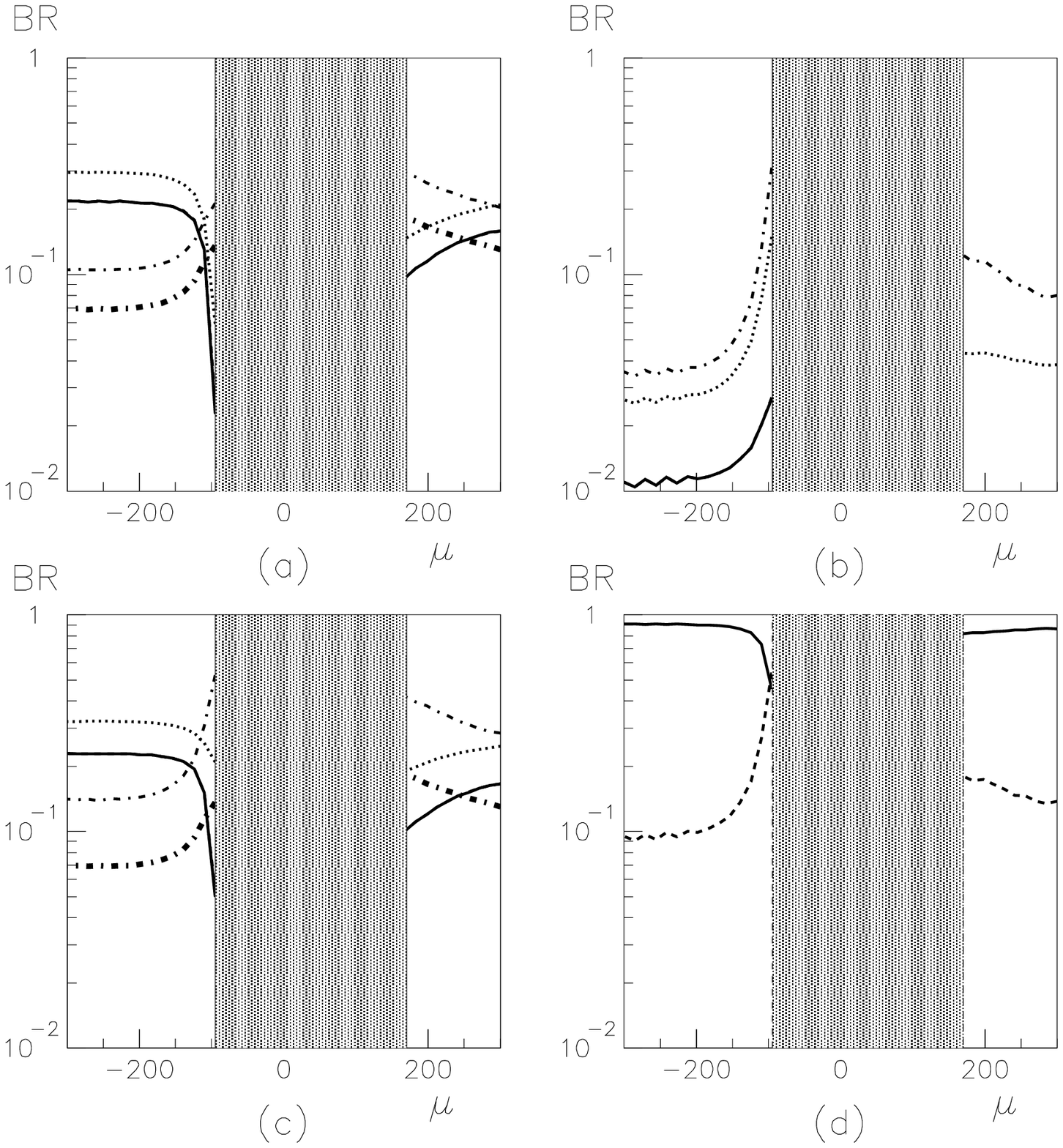}
\caption{}
\label{fig6}
\end{figure}

\begin{figure}[ht]
\hspace*{-1.0 truein}
\psfig{figure=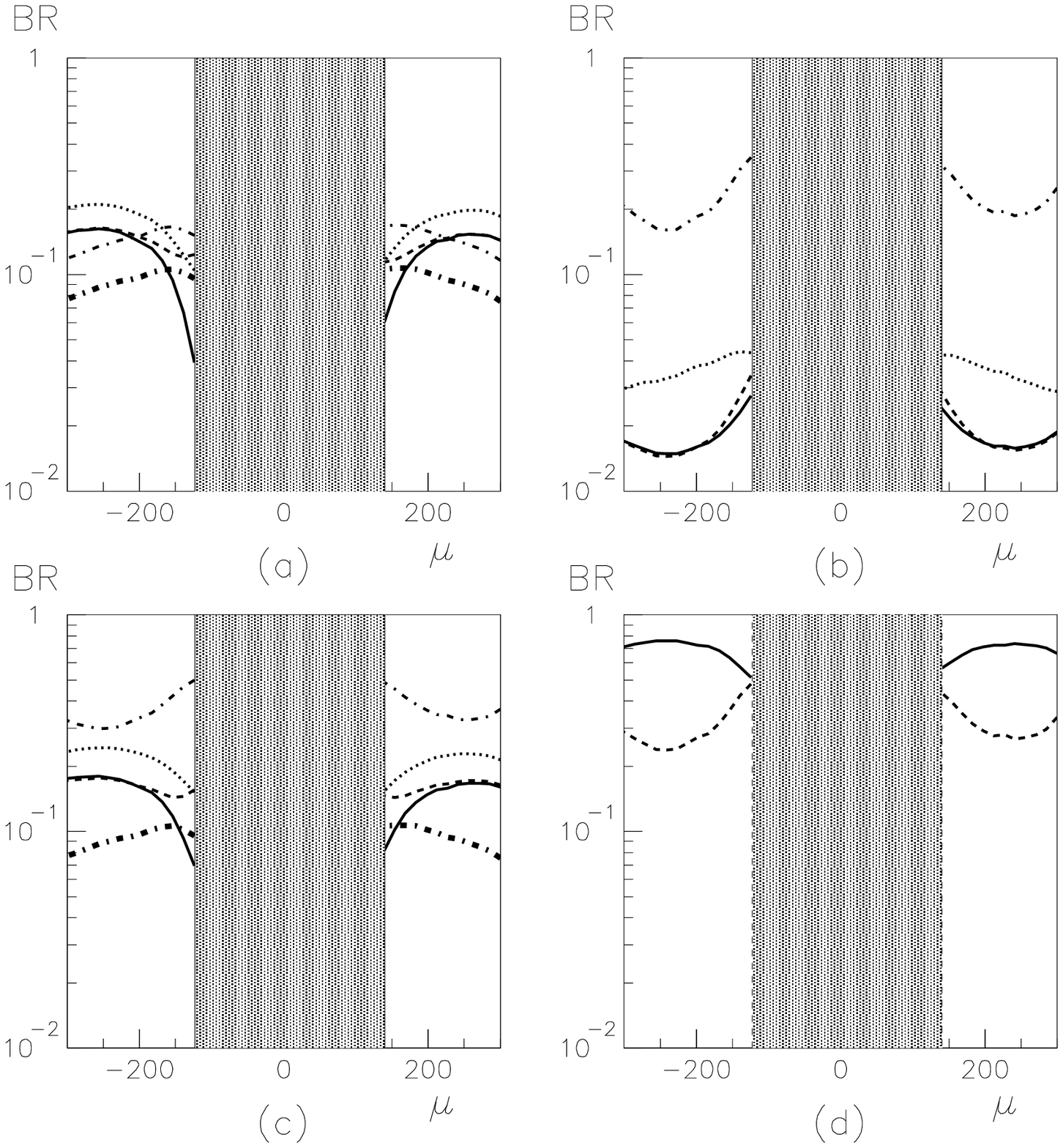}
\caption{}
\label{fig7}
\end{figure}

\begin{figure}[ht]
\hspace*{-1.0 truein}
\psfig{figure=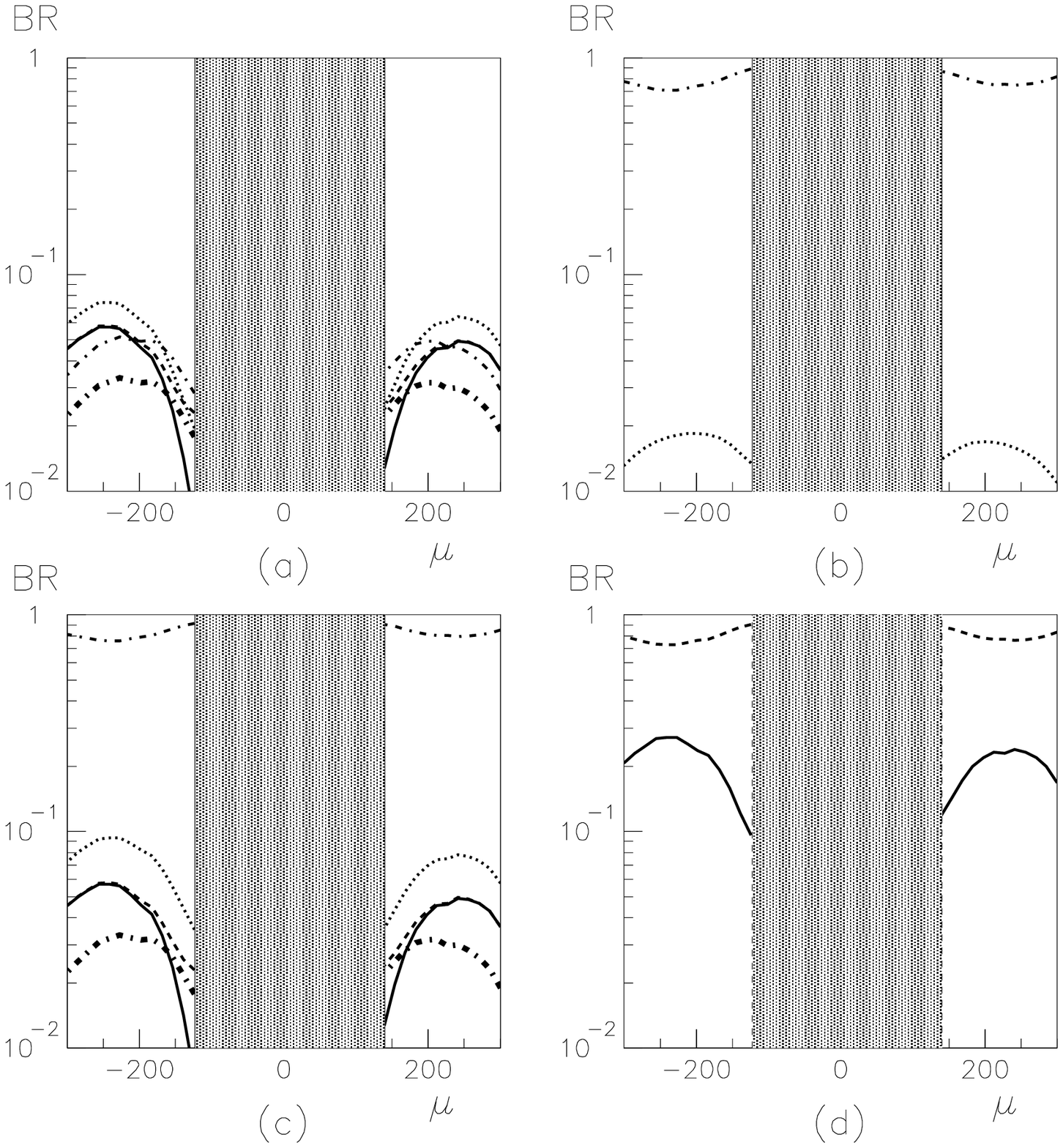}
\caption{}
\label{fig8}
\end{figure}

\end{document}